# Predictions of the hydrodynamic interpretation of quantum mechanics compared with quantum electrodynamics for low energy bremsstrahlung

MARK P. DAVIDSON[a]

[a]Spectel Research Corp., 807 Rorke Way, Palo Alto, CA 94303, USA
Email: mdavid@spectelresearch.com

ABSTRACT. It is shown that the hydrodynamic interpretation of a charged quantum particle leads to a different theoretical prediction for low energy bremsstrahlung than does quantum electrodynamics (QED). In the calculations, the electromagnetic fields are treated classically in the hydrodynamic case, but are quantized in QED. Calculations show the hydrodynamic model to have a different and more sensitive dependence on the size and shape of the radiating particle's wave packet then does QED. In particular it is shown that bremsstrahlung is sometimes greatly reduced when the force acting on the particle is localized to a volume small compared to the particle's wave packet. QED exhibits no such reduction. Therefore it is possible to test this effect experimentally.

An experiment is proposed. It involves an electron microscope with a Wien filter for producing monochromatic beam electrons and an accurate energy measurement of the particle after passing through a local force field.

## 1 Introduction

In the hydrodynamic interpretation of quantum mechanics[1-15] and the related self-field electromagnetic model[16] the probability density or current for a charged particle is interpreted as a real classical fluid density or current. The fluid's motion is described by Schrödinger's equation and it has been believed that the hydrodynamic model is equivalent to more standard interpretations. But if the charge is really interpreted as a fluid, then in the limit of soft photon emission it must couple to the classical electromagnetic field as a classical charged current. This leads to the radiation theory presented here. It is in opposition to conventional radiation theory because one conclusion that can be drawn from it is that bremsstrahlung will be suppressed in situations where the force field acts only on part of the wave packet at a given instant in time as shown for example in Figure 1. When the same problem is tackled using conventional radiation theory, it is found that no suppression occurs.

Bremsstrahlung and synchrotron radiation have been studied extensively in the physics literature[17-27]. This paper deals only with non-relativistic particles. The relativistic case is more difficult to study experimentally because it is harder to prepare very long wave packets. The particle beam will be assumed to have a low enough flux that only one particle at a time is scattering off of the force so that the particles in the beam may be considered isolated.

First an analysis of a hydrodynamic model is presented. Then a detailed radiation calculation based on conventional non-relativistic radiation theory is given. Finally an experiment is proposed to test which theory is correct.

## 2  Radiation from a charged fluid in the hydrodynamic model

In this analysis the charged current density for the Schrödinger equation is treated as a classical current. This is a very plausible assumption in the soft photon limit of a weak force field if one accepts a hydrodynamic picture. It is hard to imagine what else could be an alternative. Assume a purely electrostatic force causing the acceleration. Start with Schrödinger's equation for a charged particle ignoring spin

$$\left[ \frac{1}{2M} \mathbf{P}^2 + q\Phi_e \right] \Psi = i\hbar \frac{\partial \Psi}{\partial t}, \quad \mathbf{p} = -i\hbar\nabla \tag{1}$$

The charge and current densities are

$$\rho(\mathbf{x}, t) = q\Psi^*\Psi; \quad \mathbf{J}(\mathbf{x}, t) = \frac{q\hbar}{2Mi} \left\{ \Psi^*\nabla\Psi - \Psi\nabla\Psi^* \right\} \tag{2}$$

$$J^\mu = (c\rho, \mathbf{J}), \quad x^\mu = (ct, \mathbf{x}) \tag{3}$$

Now consider the electromagnetic field generated by these sources.

$$\partial_\mu F^{\mu\nu} = \frac{4\pi}{c} J^\nu; \quad F^{\mu\nu} = \partial^\mu A^\nu - \partial^\nu A^\mu \tag{4}$$

$$A^\mu = (\Phi, \mathbf{A}) \tag{5}$$

(the metric is timelike, ie. $g^{00} = 1$). The fields are determined up to addition of an arbitrary free field, but this free field will not contribute to any radiation, and so it can be chosen for convenience. Working in the Lorentz gauge ($\partial^\mu A_\mu = 0$) one then has the classical result

$$A^\mu(\mathbf{x}, t) = \frac{1}{c} \int \frac{J_\mu(\mathbf{x}', t - \frac{R}{c})}{R} d^3x'; \quad R = |\mathbf{x} - \mathbf{x}'| \tag{6}$$

To be partly cognizant of relativity limits we might impose the constraint on the momentum distribution

$$\rho(\mathbf{p}) = \Psi^*(\mathbf{p})\Psi(\mathbf{p}) = 0, \text{ for } \left|\frac{\mathbf{p}}{M}\right| > (1 - \delta)c \text{ for some } \delta > 0 \tag{7}$$

With this constraint, the phase velocities of the waves making up the wave function would then have a maximum absolute value of

$$v_{max} = (1 - \delta)c < c \tag{8}$$

It is still possible to localize the Schrödinger particle with this constraint on the phase velocity, and all the higher moments calculated below may be finite provided the wave function is infinitely differentiable as a function of $\mathbf{p}$. The magnetic field is derived from the vector potential by

$$\mathbf{B}(\mathbf{x}, t) = \nabla \times \mathbf{A} = \nabla \times \frac{1}{c} \int \frac{\mathbf{J}(\mathbf{x'}, t - \frac{R}{c})}{R} d^3 x' \tag{9}$$

Define a unit vector

$$\hat{\mathbf{n}} = \frac{\mathbf{x} - \mathbf{x'}}{|\mathbf{x} - \mathbf{x'}|} \tag{10}$$

The curl can now be evaluated as follows

$$\mathbf{B}(\mathbf{x}, t) = \frac{1}{c} \int \hat{\mathbf{n}} \times \frac{\partial}{\partial R} \frac{\mathbf{J}(\mathbf{x'}, t - \frac{R}{c})}{R} d^3 x' \tag{11}$$

$$\mathbf{B}(\mathbf{x}, t) = \frac{1}{c} \int \hat{\mathbf{n}} \times \left[ -\frac{1}{R^2} \mathbf{J}(\mathbf{x'}, t - R/c) - \frac{1}{Rc} \dot{\mathbf{J}}(\mathbf{x'}, t - R/c) \right] d^3 x' \tag{12}$$

In evaluating the radiation emitted, the limit where $|\mathbf{x}| \to \infty$ is taken, and therefore the leading behavior of $\mathbf{B}$ is all that need be kept.

$$R = \sqrt{\mathbf{x}^2 + \mathbf{x'}^2 - 2\mathbf{x} \cdot \mathbf{x'}} \tag{13}$$

Define

$$R_0 = |\mathbf{x}| \tag{14}$$

And so to leading order in $R_0$

$$\hat{n} = \frac{\mathbf{x}}{|\mathbf{x}|} \tag{15}$$

$$R = R_0 - \hat{\mathbf{n}} \cdot \mathbf{x'} \tag{16}$$

And so to leading order in $R_0$

$$\mathbf{B}(\mathbf{x},t) = -\hat{\mathbf{n}} \times \frac{1}{c^2 R_0} \int \dot{\mathbf{J}}(\mathbf{x'}, t - \frac{R_0 - \hat{\mathbf{n}} \cdot \mathbf{x'}}{c}) d^3 x' \tag{17}$$

Now expand in a Taylor series. It is assumed that $\mathbf{J}$ is infinitely differentiable as a function of time and that the Taylor series converges

$$\mathbf{B}(\mathbf{x},t) =$$
$$-\hat{\mathbf{n}} \times \frac{1}{c^2 R_0} \sum_{m=1}^{\infty} \int \left( \frac{\partial^m}{\partial t^m} \mathbf{J}(\mathbf{x'}, t - \frac{R_0}{c}) \right) \left( \frac{\hat{\mathbf{n}} \cdot \mathbf{x'}}{c} \right)^{m-1} /(m-1)! \, d^3 x' \tag{18}$$

Now insert the Schrödinger current (2) into this equation, and assume that the order of summation and integration can be interchanged. One must evaluate the following integrals

$$\mathbf{I}_m(t_0) = \int \mathbf{J}(\mathbf{x'}, t_0) \left( \frac{\hat{\mathbf{n}} \cdot \mathbf{x'}}{c} \right)^{m-1} d^3 x', \text{ where } t_0 = t - R_0 / c \tag{19}$$

In terms of which $\mathbf{B}$ may be written to leading order in $R_0$

$$\mathbf{B}(x,t) = -\hat{\mathbf{n}} \times \frac{1}{c^2 R_0} \sum_{m=1}^{\infty} \frac{\partial^m}{\partial t_0^m} \mathbf{I}_m(t_0) \left( \frac{1}{c} \right)^{m-1} /(m-1)! \tag{20}$$

$\mathbf{I_m}$ takes the form

$$\mathbf{I}_m(t_0) =$$
$$\frac{q\hbar}{2Mi} \int (\hat{\mathbf{n}} \cdot \mathbf{x'})^{m-1} \left\{ \Psi^*(\mathbf{x'}, t_0) \nabla \Psi(\mathbf{x'}, t_0) - \Psi(\mathbf{x'}, t_0) \nabla \Psi^*(\mathbf{x'}, t_0) \right\} d^3 x' \tag{21}$$

this may be written as

$$\mathbf{I}_m = \frac{q}{2M} \int \Psi^*(\mathbf{x'},t_0) \left\{ \mathbf{P}\left(\hat{\mathbf{n}} \cdot \mathbf{x'}\right)^{m-1} + \left(\hat{\mathbf{n}} \cdot \mathbf{x'}\right)^{m-1} \mathbf{P} \right\} \Psi(\mathbf{x'},t_0) d^3 x' \tag{22}$$

where

$$\mathbf{P} = -i\hbar\nabla \tag{23}$$

Now we transform (22) by transforming to the Heisenberg representation.

$$\mathbf{I}_m(t_0) = \frac{q}{2M} \times$$
$$\int \Psi^*(\mathbf{x},0) \left\{ \mathbf{P}(t_0)\left(\hat{\mathbf{n}} \cdot \mathbf{X}(t_0)\right)^{m-1} + \left(\hat{\mathbf{n}} \cdot \mathbf{X}(t_0)\right)^{m-1} \mathbf{P}(t_0) \right\} \Psi(\mathbf{x},0) d^3 x \tag{24}$$

For a free particle it follows that the radiation from all the terms vanishes [28]. The term that generates Larmor radiation corresponds to $m=1$ when there are forces acting.

$$\mathbf{I}_1 = \int \Psi^*(\mathbf{x},0) \left( \frac{q}{M} \mathbf{P}(t_0) \right) \Psi(\mathbf{x},0) d^3 x$$
$$= q \int \Psi^*(\mathbf{x},0) \dot{\mathbf{X}}(t_0) \Psi(\mathbf{x},0) d^3 x \tag{25}$$

Substitute this into (20) to obtain

$$\mathbf{B}(x,t) = -\frac{q}{c^2 R_0} \hat{\mathbf{n}} \times \int \Psi^*(\mathbf{x},0) \ddot{\mathbf{X}}(t_0) \Psi(\mathbf{x},0) d^3 x$$
$$= -\frac{q}{c^2 R_0} \hat{\mathbf{n}} \times \left\langle \Psi | \mathbf{a} | \Psi \right\rangle \tag{26}$$

Where $\left\langle \Psi | \mathbf{a} | \Psi \right\rangle$ denotes the expectation value of the acceleration calculated at the retarded time $t_0$.

$$\left\langle \Psi | \mathbf{a} | \Psi \right\rangle = \int \Psi^*(\mathbf{x},0) \ddot{\mathbf{X}}(t_0) \Psi(\mathbf{x},0) d^3 x$$
$$= \frac{q}{M} \int \Psi^*(\mathbf{x},0) \left[ -\nabla \Phi_c(\mathbf{x}(t_0)) \right] \Psi(\mathbf{x},0) d^3 x$$
$$= \frac{q}{M} \int \Psi^*(\mathbf{x},t_0) \left[ -\nabla \Phi_c(\mathbf{x}) \right] \Psi(\mathbf{x},t_0) d^3 x \tag{27}$$

The electric field in a region which is far from any charges or currents can then be calculated by

$$\mathbf{E} = -\hat{\mathbf{n}} \times \mathbf{B} \tag{28}$$

The Poynting vector then is

$$S = \frac{c}{4\pi} \mathbf{E} \times \mathbf{B} = \frac{q^2}{4\pi c^3 R_0^2} \langle \Psi | \mathbf{a} | \Psi \rangle^2 \sin^2(\theta) \hat{\mathbf{n}} \qquad (29)$$

where $\theta$ is the angle between $\langle \Psi | \mathbf{a} | \Psi \rangle$ and $\hat{\mathbf{n}}$. And the total power radiated is

$$P_R = \int_0^\pi \frac{q^2}{4\pi c^3 R_0^2} \langle \Psi | \mathbf{a} | \Psi \rangle^2 \sin^2(\theta) R_0^2 \, 2\pi \sin(\theta) d\theta \qquad (30)$$

Which simplifies to

$$P_R = \frac{2}{3} \frac{q^2}{c^3} \langle \Psi | \mathbf{a} | \Psi \rangle^2 \qquad (31)$$

And the total energy radiated will be an integral over time. If the scattering in Figure 1 occurs between times 0 and T then the total energy will be

$$E_{rad} = \frac{2}{3} \frac{q^2}{c^3} \int_0^T \left| \langle \Psi | \mathbf{a}(t') | \Psi \rangle \right|^2 dt' \qquad (32)$$

This is similar to Larmor's formula, familiar from classical electromagnetism. The expectation value of the acceleration $\langle \Psi | \mathbf{a} | \Psi \rangle$ in (31) involves an integral over the force acting on the particle. If the particle has an extended wave packet, then the results can differ substantially from the classical result for a point particle. For example, consider the situation in Figure 1, where the Schrödinger wave packet is moving in the x direction and is elongated in the x direction, ie. the x momentum uncertainty is very small. Let the force field which this particle is moving through be localized in a volume whose x dimension $\delta$ is much smaller than the wave packet length L.

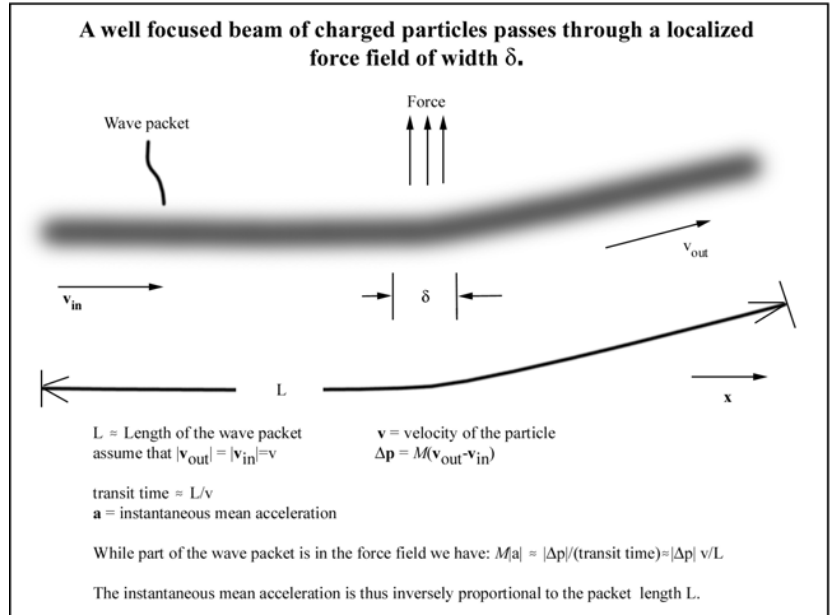

Figure 1 Localized force acting on an extended wave packet

Imagine that the wave packet in figure 1 is essentially invariant in shape along its length. We see that the instantaneous mean acceleration gets smaller inversely proportional to the wave packet's length while the impulse to the particle remains constant. The force is active on the wave packet for a time equal to the transit time, and so the total energy is the power integrated over this time. The total energy radiated, using the larmor formula, will then be, for values of L which are large compared with $\delta$

$$\text{Total Energy radiated} \approx \frac{2}{3}\frac{q^2}{c^3}\left|\left\langle\Psi\left|\mathbf{a}\right|\Psi\right\rangle\right|^2 (\text{L/v})$$

$$\approx \frac{2}{3}\frac{q^2}{c^3}\left(\frac{\Delta p \cdot \mathbf{v}}{M \cdot \text{L}}\right)^2 (\text{L/v})$$

(33)

This expression goes to zero as 1/L in the limit of large L. Even though the total impulse imparted to the particle by the force is held constant, the total radiated energy goes to zero for very long wave packets. This is the effect that is being predicted here if the hydrodynamic picture is correct. It says that by preparing very long wave packets, bremsstrahlung can be suppressed and a particle can be accelerated with greatly reduced radiation if the hydrodynamic interpretation of quantum mechanics correctly describes nature.

## 3 A treatment of the same problem using the conventional non-relativistic quantum theory of radiation

Again we assume that the beam current is low enough that only a single particle at a time is interacting with the external potential field. We must analyze the bremsstrahlung while paying careful attention to the wave packet dependence. Goldberger and Watson give a rather rigorous treatment of wave packets in scattering theory[29]. The more common textbook discussions of quantum scattering theory ignore the localized wave packet nature of the initial state and formu-late the scattering in terms of energy eigenstates for the initial and final state. It is simpler to do this because the energy eigenstates can then taken to be ortho-normal and complete, and therefore any physical wave packet can be built up from them so there is no loss of generality. However, ignoring the wave packet obscures certain effects, such as the one we are discussing here.

We work from the treatment in[18], where the radiative transition matrix for the emission of a single photon of is calculated to be (ibid 57.27):

$$\langle \beta | H'(t) | \alpha \rangle = \langle \beta | T_B | \alpha \rangle \exp(i \frac{E_\beta - E_\alpha + \hbar\omega}{\hbar} t) \tag{34}$$

We have digressed slightly from the notation (ibid) in that $\alpha$ and $\beta$ denote the initial and final states of the charged particle alone as well as denoting the complete initial and final states including photons. Here $E_\alpha$ is the initial charged particle energy and $E_\beta$ it's final energy, and where

$$\langle \beta | T_B | \alpha \rangle = \frac{iq\hbar}{McL^{3/2}} \left( \frac{2\pi\hbar c(n_{k\lambda}+1)}{k} \right)^{1/2} \times$$
$$\int u_\beta^*(\mathbf{r}) \exp(-i\mathbf{k}\cdot\mathbf{r}) \boldsymbol{\varepsilon}_{k\lambda} \cdot \nabla u_\alpha(\mathbf{r}) d^3r \tag{35}$$

In this formula, only the creation operator part of the perturbing Hamiltonian (ibid 57.21) is included, and this results in the operator being non-Hermitian in (35). Note that because $\mathbf{k}$ and $\boldsymbol{\varepsilon}$ are orthogonal, the order of the exponential term and the gradient does not matter. In this expression $u_\beta$ is the unperturbed, normalized, and time-independent final state energy eigenfunction for the charged particle, and $u_\alpha$ is the unperturbed initial state energy eigenfunction, and $n_{k\lambda}$ is the number of photons present of the type being radiated in the initial state. We shall set $n_{k\lambda}$ to zero for the remainder of this treatment. Here we mean unperturbed in the sense that the effect of the radiative process has not been included, but the effect of the external potential in the single particle Schrödinger equation has been fully included. Periodic box boundary conditions have been imposed in a box of side L on all the fields.

The eigenfunctions satisfy the stationary equation where it is assumed that the external potential does not depend on time.

$$\left[ \frac{1}{2M} \mathbf{P}^2 + q\Phi_c \right] u_\alpha = E_\alpha u_\alpha \tag{36}$$

We may rewrite (34) as

$$\langle \beta | H'(t) | \alpha \rangle =$$
$$\frac{-q\, e^{i\omega t}}{McL^{3/2}} \left( \frac{2\pi\hbar c}{k} \right)^{1/2} \int u_\beta^*(\mathbf{x}) \exp(-i\mathbf{k}\cdot\mathbf{x}(t)) \boldsymbol{\varepsilon}_{k\lambda} \cdot \mathbf{P}(t) u_\alpha(\mathbf{x}) d^3r \tag{37}$$

In the situation of interest here, the initial state is not an energy eigenstate, but rather is initially a wave packet – a plane wave modulated by an envelope function. This wave packet ψ can be written as a superposition of energy eigenstates

$$\Psi(\mathbf{x},t) = \sum_{\alpha} c_{\alpha} e^{-iE_{\alpha}t/\hbar} u_{\alpha}(\mathbf{x}) \tag{38}$$

Therefore, the transition amplitude for the emission of a single photon with the charged particle ending in a particular state is given by

$$\langle \beta | H'(t) | \Psi \rangle =$$
$$\frac{-q \, e^{i\omega t}}{McL^{3/2}} \left( \frac{2\pi\hbar c}{k} \right)^{1/2} \int u_{\beta}^{*}(\mathbf{x}) \exp(-i\mathbf{k}\cdot\mathbf{x}(t)) \boldsymbol{\varepsilon}_{\mathbf{k}\lambda} \cdot \mathbf{P}(t)\Psi(\mathbf{x},0)d^{3}r \tag{39}$$

Physically, this transition matrix element will start off being zero before the wave packet overlaps the force field, then the scattering will occur, and finally it will end up zero again after the wave packet has entirely passed through the force field as is described for example in[30] chapter 10-d.

Assuming that the particle is released at time t = 0, and is initially far away from the force field, the total transition amplitude for scattering into final state β by time T is

$$A_{\beta}(T) = \int_{0}^{T} \langle \beta | H'(t) | \Psi \rangle dt \tag{40}$$

If T is large enough so that the wave packet has already passed through the scattering region by this time, then further increasing T should make no difference in the result as all of the scattering that is going to occur has already occurred. The probability of photon emission and the energy radiated in the time interval 0 to T are

$$P = \sum_{\beta} \sum_{\mathbf{k}\lambda} \frac{1}{\hbar^{2}} \left| A_{\beta}(T) \right|^{2} \tag{41}$$

The total energy radiated is

$$E_{rad} = \sum_{\beta} \sum_{\mathbf{k}\lambda} \frac{\omega}{\hbar} \left| A_{\beta}(T) \right|^{2} \tag{42}$$

This may be written

$$E_{rad} = \sum_{\beta} \sum_{\mathbf{k}\lambda} \frac{\omega}{\hbar} \int_{0}^{T} dt' \int_{0}^{T} dt \langle \Psi | H'(t')^{\dagger} | \beta \rangle \langle \beta | H'(t) | \Psi \rangle \tag{43}$$

Now owing to the completeness of the energy eigenstates, this becomes

$$E_{rad} = \sum_{\mathbf{k}\lambda} \frac{\omega}{\hbar} \int_0^T dt' \int_0^T dt \left\langle \Psi \left| H'(t')^\dagger H'(t) \right| \Psi \right\rangle \tag{44}$$

Where

$$\left\langle \Psi \left| H'(t')^\dagger H'(t) \right| \Psi \right\rangle = e^{i\omega(t-t')} \frac{q^2}{M^2 c^2 L^3} \frac{2\pi\hbar c}{k} \times$$
$$\int \Psi^*(\mathbf{x}) e^{i\mathbf{k}\cdot\mathbf{x}(t')} \boldsymbol{\varepsilon}_{\mathbf{k}\lambda}^* \cdot \mathbf{P}(t') e^{-i\mathbf{k}\cdot\mathbf{x}(t)} \boldsymbol{\varepsilon}_{\mathbf{k}\lambda} \cdot \mathbf{P}(t) \Psi(\mathbf{x},0) d^3x \tag{45}$$

Following[18] we introduce the density of photon states and replace the summation by an integral

$$\sum_{\mathbf{k}\lambda} \rightarrow \sum_\lambda \int \frac{L^3 \omega^2}{8\pi^3 c^3} \sin(\theta) d\theta d\phi d\omega \tag{46}$$

To obtain

$$E_{rad} = \frac{q^2}{4\pi^2 M^2 c^3} \times$$
$$\sum_\lambda \int_0^T dt' \int_0^T dt \int_0^\pi d\theta \sin(\theta) \int_0^{2\pi} d\phi \int_0^\infty d\omega \omega^2 e^{i\omega(t-t')} \times$$
$$\int \Psi^*(\mathbf{x},0) e^{i\mathbf{k}\cdot\mathbf{x}(t')} \boldsymbol{\varepsilon}_{\mathbf{k}\lambda}^* \cdot \mathbf{P}(t') e^{-i\mathbf{k}\cdot\mathbf{x}(t)} \boldsymbol{\varepsilon}_{\mathbf{k}\lambda} \cdot \mathbf{P}(t) \Psi(\mathbf{x},0) d^3x \tag{47}$$

We now use the result:

$$\sum_\lambda \varepsilon_{\mathbf{k}\lambda,i}^* \varepsilon_{\mathbf{k}\lambda,j} = \delta_{i,j} - \frac{k_i k_j}{k^2} \tag{48}$$

If the k dependence of the exponential term is ignored the $\omega$ integral then has the form

$$\int d\omega \omega^2 \int_0^T dt \int_0^T dt' f^*(t') f(t) e^{i\omega(t-t')}$$
$$\approx \pi \int_0^T dt \int_0^T dt' f^*(t') f(t) \frac{\partial}{\partial t} \frac{\partial}{\partial t'} \delta(t-t')$$
$$= \pi \int_0^T dt \left| \dot{f}(t) \right|^2 \tag{49}$$

This is valid if the wavelength of the emitted radiation is much larger than the dimensions over which the force field is nonzero, ie. the soft photon limit. And therefore

$$E_{rad} = \frac{q^2}{4\pi^2 M^2 c^3} \sum_{\lambda} \pi \int\limits_0^T dt \int\limits_0^\pi d\theta \sin(\theta) \int\limits_0^{2\pi} d\phi \times$$
$$\int \Psi^*(\mathbf{x},0) \boldsymbol{\varepsilon}_{\mathbf{k}\lambda}^* \cdot \dot{\mathbf{P}}(t) \boldsymbol{\varepsilon}_{\mathbf{k}\lambda} \cdot \dot{\mathbf{P}}(t) \Psi(\mathbf{x},0) d^3 x \tag{50}$$

$$E_{rad} = \frac{q^2}{4\pi M^2 c^3} \int\limits_0^T dt \int\limits_0^\pi d\theta \sin(\theta) \int\limits_0^{2\pi} d\phi \times$$
$$\int \Psi^*(\mathbf{x},0) \left( \dot{\mathbf{P}}(t) \cdot \dot{\mathbf{P}}(t) - \frac{(\dot{\mathbf{P}}(t) \cdot \mathbf{k})^2}{k^2} \right) \Psi(\mathbf{x},0) d^3 x \tag{51}$$

$$E_{rad} = \frac{2q^2}{3M^2 c^3} \int\limits_0^T dt \times$$
$$\int \Psi^*(\mathbf{x},0) \dot{\mathbf{P}}(t) \cdot \dot{\mathbf{P}}(t) \Psi(\mathbf{x},0) d^3 x \tag{52}$$

$$E_{rad} = \frac{2q^2}{3c^3} \int\limits_0^T \left\langle \Psi \left| \mathbf{a}(t)^2 \right| \Psi \right\rangle dt' \tag{53}$$

$$\left\langle \Psi \left| \mathbf{a}(t')^2 \right| \Psi \right\rangle = \frac{1}{M^2} \int \Psi^*(\mathbf{x},0) \dot{\mathbf{P}}(t') \cdot \dot{\mathbf{P}}(t') \Psi(\mathbf{x},0) d^3 x$$
$$= \frac{1}{M^2} \int \Psi^*(\mathbf{x},t')(-q\nabla\Phi_c(\mathbf{x}))^2 \Psi(\mathbf{x},t') d^3 x \tag{54}$$

Compare this result with (32). The expression $\mathbf{a}^2$ appears inside the expectation here. Consider Figure 1 again. The total radiated energy will be independent of L because the transit time is proportional to L/v but the radiated power goes as 1/L and so the L dependence cancels out. $E_{rad}$ will be independent of L in the conventional radiation theory.

## 4 Experimental apparatus

In order to determine which theory is correct, an experiment must be capable of preparing a large wave packet beam and then measuring a small energy loss due to Bremsstrahlung. An electron microscope seems well suited for this task. Figure 2 illustrates a possible system. The most suitable electron gun would be a cold field emitter [31-33] as these have a lower energy spread than either thermal emitters, Schottky field emitters, or thermal field emitters. After the electron gun, it is desirable to have a well-collimated beam, and this is achieved by means of a collimating lens shown as a magnetic lens in Figure 2. Following the collimating lens is a so-called Wien filter which consists of crossed electric and magnetic fields "tuned" to a particular velocity at which the electric force on the charge exactly cancels the magnetic force and the desired velocity goes straight through the Wien filter with no deflection. If the velocity is greater/less than the selected velocity, then the magnetic force will be stronger/weaker than the electric force and the particle will be deflected. All but the desired velocity are eliminated from the beam by means of a pinhole aperture. The purpose of the collimating lens is to make the beam entering the Wien filter as parallel as possible so as to make the filter more effective. The wave packet length L is related to the energy spread by the Heisenberg uncertainty principle

$$L \approx \frac{\hbar \mathrm{v}}{2\Delta E} \tag{55}$$

After the beam has passed through the Wien filter it passes through an objective lens that focuses the beam. If $\alpha$ is the half-angle of the cone of rays coming from the objective lens, then the diffraction limited Rayleigh resolution (R) and depth of focus (DOF) are

$$R = 0.61 \frac{\lambda_{DB}}{\sin(\alpha)}; \quad DOF \approx \frac{\lambda_{DB}}{\sin^2(\alpha)}; \quad \lambda_{DB} = \frac{h}{p} \tag{56}$$

It is necessary to focus the beam so that it is well localized in the transverse direction. The cone half-angle $\alpha$ must be chosen so that the Rayleigh resolution is less than the dimension of the force field perpendicular to the optical axis, while at the same time the depth of focus is still large compared to the dimension of the force field along the optical axis. Usually in electron microscopy $\alpha$ is much less than 1 and a depth of focus of 10's of microns is easily achievable with a resolution of 10 nanometers. Therefore if the force field extends over a region about a micron in dimension there should be no problem is finding a suitable value for $\alpha$.

When the particle beam passes through the focus plane of the objective lens it passes between two electrodes with sharp tips. These electrodes could be sharp Tungsten needles as are used in field emitter guns or in scanning tunneling microscopy. Such needles have tips several nanometers in width. Depending on how close the tips are to one another, the spatial extent of the force field can be controlled. As the beam passes through this region it is deflected and it also radiates bremsstrahlung radiation. As a result of this the energy of the electron is reduced.

Finally the electron enters a precision energy detector. Such detectors are used in electron energy loss spectroscopy or EELS[34-36]. It must have sufficient precision to measure the energy lost to radiation.

It would also be desirable to allow conventional scanning electron microscope imaging to take place in the microscope of figure 2 so as to allow the positioning of the electron beam precisely between the two electrodes. This would be achieved by adding a conventional beam scanning yoke or electrostatic deflector to the optical system of figure 2 along with a secondary electron detector and raster scan electronics. It might also be desirable to add a photon detector to be placed near to the region of the electrodes to detect emitted radiation directly.

If the hydrodynamic model is correct the energy loss should be proportional to 1/L for large L. If the conventional quantum theory of radiation is correct then the energy loss should be independent of L for large L.

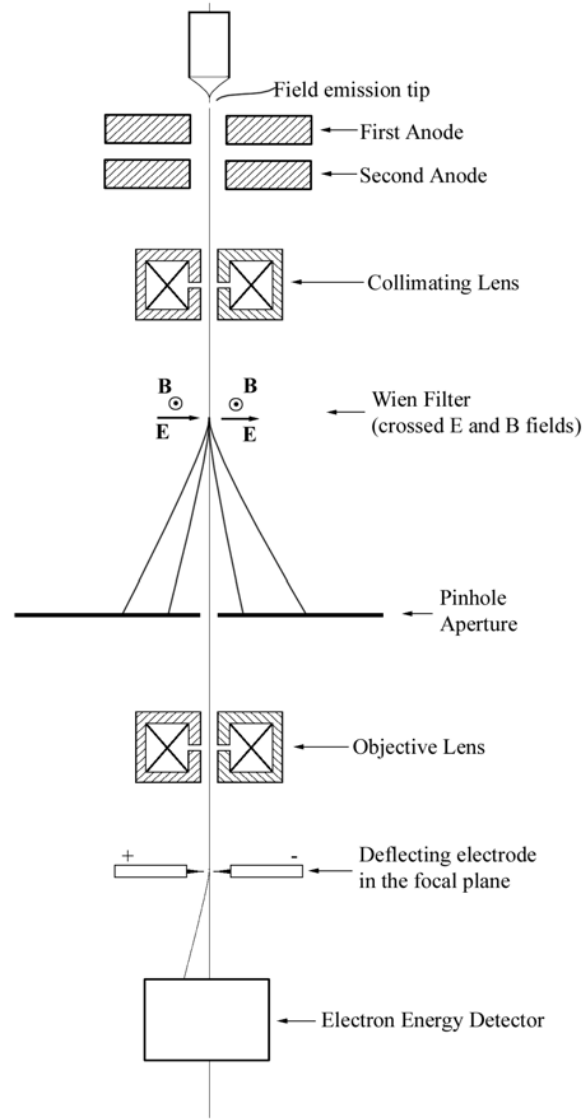

Figure 2. Apparatus for measuring Bremsstrahlung energy loss

## 5  Conclusion

It is rare that fundamental interpretations of quantum mechanics can be tested experimentally. The hydrodynamic interpretation of the Schrödinger wave asserts that a particle's charge and current densities are proportional to the Schrödinger probability current densities. Once this is granted and classical electromagnetism is used to calculate the soft photon limit of bremsstrahlung, the results here seem to follow quite clearly. If the hydrodynamic view is correct then bremsstrahlung will be suppressed when the wave packets are large compared with the region of the applied force. This is incompatible with the conventional quantum theory of radiation and QED as shown here. Table I summarizes the two cases. The end results are quite simple considering the work needed to obtain them. In the hydrodynamic model one takes the mean of the acceleration first and then squares this mean to get the instantaneous power. In QED we must calculate instead the mean of the acceleration vector squared. This subtle difference leads to experimentally testable differences.

| | |
|---|---|
| Classical Radiation Result | $E_{rad} = \dfrac{2}{3}\dfrac{q^2}{c^3}\displaystyle\int_0^T \mathbf{a}^2(t')dt'$ |
| Hydrodynamic Model Result | $E_{rad} = \dfrac{2}{3}\dfrac{q^2}{c^3}\displaystyle\int_0^T \left|\left\langle\Psi\left|\mathbf{a}(t)\right|\Psi\right\rangle\right|^2 dt'$ |
| Conventional Quantum Radiation Result | $E_{rad} = \dfrac{2q^2}{3c^3}\displaystyle\int_0^T \left\langle\Psi\left|\mathbf{a}^2(t')\right|\Psi\right\rangle dt'$ |

Table I. Summary of results

An experiment to differentiate which formula is realized in nature has been proposed. It involves and electron microscope with a cold field emitter electron gun, a collimating lens, a very strong Wien filter for producing a narrow energy beam, an objective lens, a deflection force causing the bremsstrahlung, and finally a precision energy detector.

If the hydrodynamic interpretation is correct then the radiated energy per particle will fall off as 1/L for large L. If the conventional radiation theory is correct then this energy will not decrease for increasing L.

If the hydrodynamic interpretation is confirmed by this experiment, then a rethinking of the conventional radiation theory will be required and a possibly useful means of monitoring the wave packet length will have been discovered. If, as is more likely, the conventional radiation theory is confirmed, then the hydrodynamic interpretation of quantum mechanics can be ruled out as a possible description of nature, or at least it must invoke a non-classical interaction with the electromagnetic field even in the soft photon limit to remain viable.

The results presented here can obviously be extended to magnetic forces and to force fields which act in an arbitrary direction relative to the incident beam.